\documentclass{article}
\usepackage{amsmath,amssymb,amsthm,textcomp,color}
\usepackage[all]{xy}

\usepackage[utf8]{inputenc}
\usepackage[breaklinks]{hyperref}
\theoremstyle{plain}
\newtheorem{theorem}{Theorem}
\newtheorem{corollary}[theorem]{Corollary}
\newtheorem{proposition}[theorem]{Proposition}

\theoremstyle{definition}
\newtheorem{definition}[theorem]{Definition}

\voffset = - 1.2cm
\numberwithin{equation}{section}
\numberwithin{theorem}{section}
\input xy
\xyoption{all}

\numberwithin{equation}{section}
\numberwithin{theorem}{section}

\def\pd#1#2{\frac{\partial#1}{\partial#2}}

\begin{document}

\centerline{{\Large \bf Jacobi multipliers, non-local symmetries  
  and  nonlinear oscillators}}
\vskip 0.5cm
\centerline{J.F. Cari\~nena$^\dagger$, J. de Lucas$^\ddagger$ and M.F. Ra\~nada$^\dagger$}
\vskip 0.25cm

\centerline{$^\dagger$ Departamento de  F\'{\i}sica Te\'orica and IUMA, Universidad de
Zaragoza,}
\vskip 0.1cm
\centerline{c. Pedro Cerbuna 12, 50009 Zaragoza, Spain.}
\centerline{$^\ddagger$ Department of Mathematical Methods in Physics, University of Warsaw}
\vskip 0.10cm
\centerline{ul. Pasteura 5, 02-093, Warszawa, Poland}
\vskip 0.25cm

\begin{abstract}
Constants of motion, Lagrangians and Hamiltonians admitted by a family of relevant nonlinear oscillators
are derived using a geometric formalism.  The theory of the Jacobi last multiplier allows us to find Lagrangian descriptions and constants of the motion.
An application of the jet bundle formulation of symmetries of differential equations is presented in the second part of the paper. After a short review of the general formalism, the particular case of non-local symmetries is studied in detail by making use of an extended formalism.  The theory is related to some results previously obtained by  Krasil'shchi, Vinogradov and coworkers. Finally the existence of non-local symmetries for such two nonlinear oscillators is proved.
\end{abstract}



\section{Introduction}

The interest of differential geometric techniques in the analysis of systems of ordinary differential equations (ODEs) is undeniable. Lie point symmetries, integrating factors and their generalizations are just several examples of geometric tools which have been successfully applied in the study of systems of ODEs and their related mathematical and physical problems.

In this work, we survey the geometric theory of Jacobi multipliers \cite{Ja44a,Cl09} and non-local symmetries \cite{KKV04,KV89} to study  a family of relevant nonlinear oscillators that have been attracting some attention in recent years \cite{BGS11,Ga09,GB11,MR12,CRS04,BEHR08,BEHRR11}. For instance, it was proved that they can be understood as oscillators in manifolds of constant curvature \cite{CRSS04}, they admit compatible bi-Hamiltonian structures, and their properties can also be analyzed by means of coalgebra techniques \cite{BEHR08,BEHRR11}. Some of their properties have also been obtained by means of the so-called $\lambda$-symmetries \cite{MR12}. 
 
First, we use Jacobi multipliers \cite{Ja44a,Cl09} to go over the above-mentioned oscillators from a geometrical point of view. This allows us to obtain some of their constants of motion. We also obtain known and new Lagrangian and Hamiltonian functions for these oscillators. It is worth noting that the new derived Lagrangian functions are of a non-mechanical type, i.e. they do not possess a kinetic term given by a $2$-contravariant tensor field.

Subsequently, we review a `new method' to obtain non-local symmetries developed by Gandarias and coworkers \cite{BGS11,Ga09,GB11}. We show that their procedure is a consequence of the non-local symmetry idea formalised by Krasil'shchik and Vinogradov several years before \cite{V99,Vi89,Ca07}. Despite this, Gandarias and coworkers' applications of this method are still relevant, as they illustrate that certain systems of differential equations with no classical point symmetries can admit non-local symmetries that lead to unveil their properties. 

The study of non-local symmetries demands the use of infinite-dimensional jet bundles \cite{Krasilshchik2011}. We illustrate how this geometric approach  can easily be applied to study our family of oscillators. Indeed, the calculation of such non-local symmetries in the problem under study is very similar to the case of a finite-dimensional jet bundle $J^p\pi$. It is essentially the geometrical interpretation what differs. Additionally, our techniques provide very simple and relevant examples of finite-dimensional diffieties describing ODEs, which is not the usual approach as they are mainly concerned with infinite-dimensional manifolds describing PDEs. 

The use of infinite dimensional jet bundles provides other advantages. Many structures of $J^p\pi$, e.g. the Cartan distribution, become simpler when defined on the infinite-dimensional jet bundle $J^\infty\pi$. Moreover, we can define geometric structures on this latter space that cannot properly be defined on $J^p\pi$, e.g. the total derivative. Moreover, $J^\infty\pi$ possesses a geometric structure richer than that of $J^p\pi$, e.g.  it admits Lie symmetries than cannot be described in terms of Lie symmetries defined on $p$-order jet bundles \cite{Ca07}.

Apart from showing the usefulness of infinite-dimensional jet bundles and reviewing Gandarias' results, we devise a new idea to easily determine non-local symmetries for certain systems of higher-order differential equations. As an application, we retrieve in a natural and rigorous way a result given by Gandarias as an ansatz for the oscillators of this work \cite{BGS11}.

The paper goes as follows. Section 2 is devoted to surveying the theory of Jacobi multipliers and its relation with  Lagrangians and constants of motion. In Section 3 we apply Jacobi multipliers to study relevant types of oscillators. In Section 4, non-local symmetries of differential equations are briefly presented. We show that the method developed by
Gandarias and coworkers \cite{BGS11,Ga09} reduces to  the non-local symmetry concept developed by Krasil'hinski and Vinogradov and we apply this idea to the mentioned oscillators in Section 5. A method to improve the derivation of such non-local symmetries is provided in Section 6 and we summarise our results and comment on our future work in Section 7.

\section{Jacobi multipliers, Lagrangians and constants of motion}

Let $M$ stand for an oriented  manifold, i.e. $M$ is  equipped with a volume form $\Omega$. Given a vector field $X$ on $M$, we call {\it divergence} of $X$ relative to $\Omega$ the unique function ${\rm div}\, X:M\rightarrow\mathbb{R}$ satisfying \cite{Ja44a,Cl09}:
$$
\mathcal{L}_{X}\Omega=({\rm div} X)\,\Omega\,.
$$
A {\it Jacobi multiplier} for $X$ is a non-vanishing function $\mu:M\rightarrow\mathbb{R}$ satisfying that $\mu( i(X)\Omega)$ is a closed form, or equivalently $\mathcal{L}_X(\mu\,\Omega)=0$. In other words, $\mu$ is such that 
$$
{\rm div}(\mu X)=0,\qquad \mu(x)\neq 0,\qquad \forall x\in M\,.
$$
{}From  $\mathcal{L}_X(f\,\Omega)=(Xf+f\,\mathrm{div}\,X)\,\Omega$,  for every $f\in C^\infty(M)$,  we see that a function $\mu$ is a Jacobi multiplier for $X$ if and only if  $\mu$ does not vanish and satisfies
\begin{equation}
\mu \,{\rm div}X + X\mu = 0\,.\label{JLMcond}
\end{equation}
A function $I:M\rightarrow \mathbb{R}$ is a 
first-integral of $X$, i.e. $XI=0$,  if and only if ${\rm d}I$ annihilates the generalized distribution $\mathcal{D}=\{v\in {\rm T}M\,|\, \exists \,p\in M, v=X_p\}$ generated by $X$, i.e. $({\rm d}I)_p(X_p)=0$ at each $p\in M$. Let us restrict ourselves to an open subset $U=\{p\in M: X_p\neq 0\,,\, ({\rm d}I)_p\neq 0\}$ of a two-dimensional manifold $M$. In this case, $i(X)\Omega$ defines a one-dimensional codistribution annihilating $\mathcal{D}$ and hence $X$ admits a Jacobi multiplier $\mu$ such that $\mu\, i(X)\Omega={\rm d}I$. Using that $(i(X)\Omega)\wedge {\rm d}f=-(i(X){\rm d}f)\wedge \Omega=-(Xf)\Omega$, for each $f\in C^\infty(U)$, we find that 
${\rm d}I\wedge {\rm d}f=-\mu \,(Xf)\,\Omega$.
This expression also shows that $f$ is a first-integral of $X$ if and only if ${\rm d}I\wedge {\rm d}f=0$, i.e. $f$ is a function of $I$ and $f=\varphi(I)$ for a certain real function $\varphi:\mathbb{R}\rightarrow \mathbb{R}$.

Jacobi multipliers have many applications \cite{Cl09,NL08a,NT08b,NL08c,GN15,BGM14,MR14}. In particular, we are interested in their use to construct Lagrangians and constants of motion for second-order differential equations  \cite{NL08b,NT08a,MR09,CGK}. Let us briefly survey this topic.
 
Assume hereafter that ${\rm T}\mathbb{R}$ is endowed with the volume form $\Omega={\rm d}x\wedge {\rm d}v$. Consider a second-order differential equation
\begin{equation}\label{SODE}
\frac{{\rm d}^ 2x}{{\rm d}t^2}=F\left(x,\frac{{\rm d}x}{{\rm d}t}\right),\qquad x\in\mathbb{R}\,,
\end{equation}
with $F:{\rm T}\mathbb{R}\rightarrow \mathbb{R}$ being an arbitrary function. Adding a new variable $v\equiv {\rm d}x/{\rm d}t$, we see that (\ref{SODE}) can be rewritten as
\begin{equation}\label{FODE}
\left\{\begin{aligned}
\frac{{\rm d}x}{{\rm d}t}&=v\,,\\
\frac{{\rm d}v}{{\rm d}t}&=F\left(x,v\right)\,,
\end{aligned}\right.
\end{equation}
whose particular solutions are integral curves of the vector field on ${\rm T}\mathbb{R}$ given by
\begin{equation}
\Gamma=v\frac{\partial}{\partial x}+F(x,v)\frac{\partial}{\partial v}.\label{aSODEvf}
\end{equation}
A first-integral of $\Gamma$ is usually called a constant of  motion for $\Gamma$ or, equivalently, for system (\ref{FODE}). By substituting $v$ by ${\rm d}x/{\rm d}t$, this first-integral gives rise to a constant of motion to (\ref{SODE}). 
As $\textrm{div\,} \Gamma=\partial F/\partial  v$ in this case, then the Jacobi's multiplier condition (\ref{JLMcond}) amounts to 
\begin{equation}
v\pd \mu x+\pd{(\mu\,F)}v=0\,. \label{multcond1}
\end{equation}
The Jacobi multiplier $\mu$ satisfying this condition is also called a Jacobi multiplier
for the second-order differential equation (\ref{SODE}).

\begin{theorem}\label{Main1}
The differential equation determining the solutions of the
Euler--Lagrange equation defined by a regular Lagrangian function $L(x,v)$ possesses, when written in its normal form (\ref{SODE}),
a Jacobi multiplier given by the function  
\begin{equation}
\mu=\frac{\partial^2L}{\partial v^2}.\label{multL}
\end{equation}
Conversely, if  $\mu$ is a Jacobi multiplier for a second-order  differential
equation (\ref{SODE}), then (\ref{SODE}) admits a regular Lagrangian $L(x,v)$ satisfying (\ref{multL}).
\end{theorem}
\begin{proof}
Assume $L$ to be a regular Lagrangian for (\ref{SODE}) and define the non-vanishing function $\mu$ by (\ref{multL}).  Note that then $F$ is given by 
$$F(x,v)=\frac 1\mu\left(\pd Lx-v\pd{^2L}{x\partial v}\right)\,,
$$ 
and, using this,  we see that  
$$
\begin{aligned}
v\pd \mu x+\pd{}v\left(\mu F\right)&=v\pd \mu x+\pd{}v\left(\pd Lx-v\pd{^2L}{x\partial
      v}\right)=v\pd{^3L}{v^2\partial x}+\pd{^2L}{x\partial
    v}-\pd{^2L}{x\partial v}-v\pd{^3L}{v^2\partial x}=0\,.
\end{aligned}
$$
Therefore,  $\mu$  given by (\ref{multL}) satisfies the Jacobi multiplier equation (\ref{multcond1}). Since $L$ is assumed to be regular, then the function
$\mu$ given by (\ref{multL}) does not vanish and becomes a Jacobi multiplier of (\ref{SODE}). 

Conversely, if $\mu$ is  a Jacobi multiplier 
for (\ref{aSODEvf}), then the functions $L$ satisfying (\ref{multL}) are of the form
\begin{equation}
L(x,v)=\int^v {\rm d}v'\int^{v'}\mu(x,\zeta) \,{\rm d}\zeta+ \phi_1(x)\,v+\phi_2(x)\,\label{posL}
\end{equation}
for arbitrary functions $\phi_1,\phi_2:\mathbb{R}\rightarrow \mathbb{R}$. The term $\phi_1(x)\,v$ is a gauge term which 
can be fixed equal to zero, i.e. $L$ can be assumed to be of the form 
\begin{equation}
L(x,v)=\int^v {\rm d}v'\int^{v'}\mu(x,\zeta) \,{\rm  d}\zeta+\phi_2(x)\,.\label{posLsingauge}
\end{equation} 
and the function $\phi_2$ can be chosen in a unique way\cite{CGR09}, up to a constant, so that the Euler--Lagrange equation reproduces the equation for the integral curves for the
given vector field (\ref{aSODEvf}). Indeed, using (\ref{posLsingauge}) we see that 
$$
\frac{\partial L}{\partial x}=\int^v{\rm d}v'\int^{v'}\frac{\partial \mu}{\partial x}(x,\zeta)\, {\rm d}\zeta+ \frac{{\rm d} \phi_2}{{\rm d} x}(x)\,,
$$ and in order to   the  Euler--Lagrange equation for the Lagrangian (\ref{posLsingauge}) to give the dynamics we should have: 
\begin{equation*}
\int^v\!\!{\rm d}v'\int^{v'}\pd{\mu}x(x,\zeta) \,
{\rm d}\zeta+\frac{{\rm d}\phi_2}{{\rm d}x}(x)=v
\int^v\pd{\mu}x(x,\zeta) \, {\rm d}\zeta+\mu(x,v)F(x,v) 
 \,.
\end{equation*}
But note that
\begin{multline}
{\displaystyle{\pd{}v}}\left(v{\displaystyle{\int^v}}\,{\displaystyle{\pd{\mu}x}}(x,\zeta) \,
  {\rm d}\zeta+\mu(x,v)F(x,v)- {\displaystyle{\int^v{\rm d}v'\int^{v'}}}\,{\displaystyle{\!\!\!\pd{\mu}x}}(x,\zeta){\rm d}\zeta\right)\\
=v{\displaystyle{\pd
  \mu x}}(x,v)+F(x,v){\displaystyle{\pd \mu v(x,v)}}+\mu(x,v){\displaystyle{\pd F v(x,v)}}\,,
\end{multline}
which vanishes because of the multiplier condition (\ref{multcond1}). Consequently, the
function  $\phi_2$ exists and is uniquely determined,  up to a constant, by
\begin{equation}
\label{phidos}
\frac{{\rm d}\phi_2}{{\rm d}x}(x)=v\int^v\,\pd{\mu}x(x,\zeta) \,
  {\rm d}\zeta+\mu(x,v)F(x,v)-\int^v{\rm d}v'\int^{v'}\,\pd{\mu}x(x,\zeta)\,{\rm d}\zeta.
\end{equation}

An integration by parts in  the double integral leads to
\begin{equation*}
\int^v{\rm d}v'\int^{v'} \pd \mu x(x,\zeta)\,{\rm d}\zeta=
\\v\int^v \pd \mu x(x,\zeta)\,{\rm d}\zeta -\int^vv'\, \pd \mu x(x,v')\,{\rm d}v',
\end{equation*}
that when substituted in (\ref{phidos}) gives
$$
\frac{{\rm d}\phi_2}{{\rm d}x}(x)=\mu(x,v)\, F(x,v)+\int^v\zeta\,\pd{\mu}x(x,\zeta)\,{\rm d}\zeta,
$$
which using the Jacobi multiplier condition (\ref{multcond1}) reduces to 
$$
\frac{{\rm d}\phi_2}{{\rm d}x}(x)=\mu(x,v)F(x,v)-\int_{v_0}^v \pd{(\mu\,F)}{\zeta}(x,\zeta)\,{\rm d}\zeta,
$$
and therefore $\phi_2$ is given by:
\begin{equation}
\phi_2(x)=\int^x (\mu\,F)(\zeta,v_0)\,{\rm d}\zeta\,.\label{phidos2}
\end{equation}
Additionally, since $\mu$ is non-vanishing and in view of (\ref{multL}), we obtain that $L$ is regular.
\end{proof}

Another remarkable  result concerning the inverse problem is given in the following Proposition \cite{Le81,Lo96a}.

\begin{proposition}\label{LagInt}
If $I$ is a constant of motion for the vector field $\Gamma$ 
at a point $\xi\in {\rm T}\mathbb{R}$ where $\Gamma_\xi\neq 0$ and $({\rm d}I)_\xi\neq 0$, then 
\begin{equation}\label{Lsol}
 L(x,v)=v\int^v\frac{I(x,\zeta)}{\zeta^2} {\rm d}\zeta
\end{equation}
 is a Lagrangian for the given vector field around a neighborhood of $\xi$.
\end{proposition}
\begin{proof}
Since $\Gamma$ and ${\rm d}I$ do not vanish at $\xi$, there exists around this point a Jacobi multiplier $\mu$ for $\Gamma$ relative to $\Omega={\rm d}x\wedge {\rm d}v$ such that 
$$ 
  \mu\, i(\Gamma)\Omega={\rm d}I\ \Longleftrightarrow\
   \mu \bigl(v\, {\rm d}v-F(x,v)\, {\rm d}x \bigr) = {\rm d}I.
$$
Therefore,
$$
 \mu=\frac{1}{v}\, \pd Iv,\qquad -\mu F=\frac{\partial I}{\partial x}. 
$$
 
In view of Theorem \ref{Main1}, there exists a Lagrangian $L$ such that 
$$\pd{^2L}{v^2}=\frac  1v \, \pd Iv,
$$
from where, an integration by parts leads to  
$$ 
\begin{aligned}
 \pd Lv&= \int^v\frac 1\zeta \pd I{\zeta}\,{\rm d}\zeta=\frac {I(x,v)}{v}+\int^v\frac{I(x,\zeta)}{\zeta^2}\,{\rm d}\zeta=\pd{}{v}\left( v\int^v\frac{I(x,\zeta)}{\zeta^2}\,{\rm d}\zeta\right).
 \end{aligned}
$$
This shows that $L$ must be given by
$$
L(x,v)=v\int^v\frac{I(x,\zeta)}{\zeta^2}{\rm d}\zeta+\phi(x)
$$
for a certain $\phi:x\in \mathbb{R}\mapsto \phi(x)\in \mathbb{R}$. Imposing $L$ to be a Lagrangian for $\Gamma$ and recalling that $\partial I/\partial x=-\mu F$, we obtain
$$
\frac{\partial L}{\partial x}-\frac{\rm d}{{\rm d}t}\frac{\partial L}{\partial v}=0\Rightarrow - v\int^v\frac{[\mu F](x,\zeta)}{\zeta^2}{\rm d}\zeta+\frac{{\rm d}\phi}{{\rm d}x}(x)-\frac{\rm d}{{\rm d}t}\left(\frac {I(x,v)}{v}+\int^v\frac{I(x,\zeta)}{\zeta^2}\,{\rm d}\zeta\right)=0.
$$
Hence,
$$
- v\int^v\frac{[\mu F](x,\zeta)}{\zeta^2}{\rm d}\zeta+\frac{{\rm d}\phi}{{\rm d}x}-\frac{\partial I}{\partial x}-\frac{F}{v}\frac{\partial I}{\partial v}+\frac{F}{v^2}I+v\int^v\frac{[\mu F](x,\zeta)}{\zeta^2}{\rm d}\zeta-\frac{FI}{v^2}=\frac{{\rm d}\phi}{{\rm d}x}=0.
$$
Hence, $\phi$ is an irrelevant constant and we obtain that  $L$ is essentially given, up to an also irrelevant gauge term, by  (\ref{Lsol}).
\end{proof}

Apart from providing a variational description for second-order differential equations (SODEs) as (\ref{SODE}), Jacobi multipliers can also be employed to derive their
$t$-independent constants of motion, namely first-integrals for the associated $\Gamma$. More specifically, given two Jacobi multipliers $\mu_1$ and $\mu_2$ of the vector field $\Gamma$, the function
$$
\varphi = \frac{\mu_1}{\mu_2}
$$
is a constant of motion for (\ref{FODE}) and, by substituting $v$ by ${\rm d}x/{\rm d}t$, we obtain a $t$-independent constant of motion for (\ref{SODE}). Indeed, as ${\rm div}(\mu_i \Gamma)=\Gamma\mu_i+\mu_i\, {\rm div}\Gamma=0$, for $i=1,2$, it follows
$$
\Gamma\varphi =\frac{\mu_2\Gamma\mu_1-\mu_1\Gamma \mu_2}{\mu_2^2}=0.
$$

Consequently,  the non-uniqueness of such a
  Lagrangian function, i.e. the existence of alternative Lagrangians 
 can be used to determine  constants of the
motion as it was proved in \cite{CS66} for the one-dimensional case and generalized in \cite{HH81} for the
multidimensional case (see also \cite{CI83}  for a geometric approach).

In addition, given a non-vanishing $t$-independent constant of motion $\varphi$ for $\Gamma$, then $\mu_1\varphi$ is a new Jacobi multiplier for $\Gamma$. This shows that given a fixed Jacobi multiplier $\mu_1$ for $\Gamma$,  any other Jacobi multiplier 
$\mu$ for $\Gamma$ arises as the product of $\mu_1$   times a non-vanishing  function $G\in C^\infty(\mathbb{R})$ of a given nontrivial constant of motion $\varphi_1$ for $\Gamma$, i.e. 
 $\mu=G(\varphi_1)\mu_1$.

\section{Jacobi multipliers and nonlinear oscillators}

Let us now use the above results to analyse the nonlinear oscillators
\begin{equation}\label{NonL1}
\frac{{\rm d}^2x}{{\rm d}t^2}-\frac{kx}{1+kx^2}\left(\frac{{\rm d}x}{{\rm d}t}\right)^2+\frac{\alpha^2x}{1+kx^2}=0,\qquad \alpha\in \mathbb{R},
\end{equation}
and
\begin{equation}\label{NonL2}
\frac{{\rm d}^2x}{{\rm d}t^2}+\frac{kx}{1+kx^2}\left(\frac{{\rm d}x}{{\rm d}t}\right)^2 
+ \frac{\alpha^2x}{(1+kx^2)^3}=0, {\qquad} \alpha\in \mathbb{R},
\end{equation}
which have recently been drawing some attention \cite{BGS11,CGK,CRSS04,ML74,BEHR08}. For instance, the Hamiltonian description of the quantum analogues of both systems led to suggest a Lagrangian description for nonlinear oscillators\cite{ML74} . Here, $x\in \mathbb{R}$ when $k\geq0$ but $|x|\neq 1/\sqrt{|k|}$ for $k<0$. For simplicity,  we study the bounded motions with $x\in (-1/\sqrt{-k},1/\sqrt{-k })$ when $k<0$.  Some generalizations of these results to higher-order dimensions
for (\ref{NonL1})   were devised in \cite{CRSS04} and some of the non-local symmetries for (\ref{NonL1}) and (\ref{NonL2}) were described in \cite{BGS11}.

The second oscillator (\ref{NonL2}) is the one-dimensional case of the Hamiltonian superintegrable system studied in \cite{BEHR08}. Apart from its superintegrability, 
this system has attracted some attention due to the fact that it can be investigated through an $\mathfrak{sl}(2,\mathbb{R})$--Poisson coalgebra (cf. \cite{BEHR08}). 
 The straightforward generalization to $n$-dimensions of both oscillators leads to oscillators of constant and variable curvature \cite{BEHR08}.

\subsection{First nonlinear oscillator }

We can write (\ref{NonL1}) as a first-order system:
\begin{equation}\label{AsoNonL}
\left\{
\begin{aligned}
\frac{{\rm d}x}{{\rm d}t}&=v,\\
\frac{{\rm d}v}{{\rm d}t}&=\frac{(kv^2-\alpha^2)x}{1+kx^2}.
\end{aligned}\right.
\end{equation}
System (\ref{AsoNonL}) describes the integral curves of the vector field
$$
\Gamma=v\frac{\partial}{\partial x}+\frac{x(kv^2-\alpha^2 )}{1+kx^2}\frac{\partial}{\partial v},
$$
and as 
$\textrm{div\,}\Gamma = 2kxv/(1+kx^2)$, 
its Jacobi multipliers, $\mu:{\rm T}\mathbb{R}\rightarrow \mathbb{R}$, are the non-vanishing solutions of the equation
$$
\textrm{div\,}(\mu \Gamma)=\Gamma\mu+\mu\,\textrm{div\,}\Gamma=0,
$$
that in this case can be written as 
\begin{equation}
v\frac{\partial \mu}{\partial x}+\frac{x(kv^2-\alpha^2)}{1+kx^2}\frac{\partial \mu}{\partial v}+\frac{2\mu kxv}{1+kx^2}=0.\label{PDEs}
\end{equation} 
This equation can explicitly be solved by the {\it method of characteristics}. This method reduces solving the above PDE to determining the integral curves of a vector field: the so-called {\it characteristic curves}. When characteristics are determined, solutions for the PDE are obtained by gluing them together giving rise to a hypersurface (see \cite{Olver,Me05} for details). The characteristic curves of (\ref{PDEs}) can be described by means of the referred to as {\it characteristic system} \cite{Olver,Me05}, namely
\begin{equation}\label{CharSys}
\frac{{\rm d}x}{v}=\frac{(1+kx^2){\rm d}v}{x(kv^2-\alpha^2)}
=-\frac{(1+kx^2){\rm d}\mu}{2 kxv\mu}.
\end{equation}
Let us solve these equations for $k=0$, i.e. the harmonic oscillator. In this case, we have ${\rm div}\,\Gamma=0$ and 
Jacobi multipliers become mere first-integrals of $\Gamma$. The characteristic curves are given by
$$
\mu =\Upsilon_1,  \qquad   x^2+v^2=\Upsilon_2,
$$
where $\Upsilon_1, \Upsilon_2$ are real constants. This means that Jacobi multipliers are non-vanishing functions of the form 
 $\mu=\mu(x^2+v^2)$. 
 
By integrating the characteristic equations for $k\neq 0$, we find that the characteristic curves are given by
$$
\mu (1+k x^2)=\Upsilon_1,  \qquad\qquad   \frac{1+kx^2}{kv^2-\alpha^2}=\Upsilon_2,
$$
where $\Upsilon_1, \Upsilon_2$ are real constants. We know that any surface $(x,y,\mu(x,v))$  obtained by gluing characteristic curves is a solution to (\ref{PDEs}), namely any subset $(x,y,\mu)$ of 
points of $\mathbb{R}^3$ satisfying
$$
K({(1+kx^2)}/{(kv^2-\alpha^2)},\mu (1+k x^2))=0,
$$ 
for a fixed function $K:\mathbb{R}^2\rightarrow \mathbb{R}$, with $\partial_2K\neq 0$ (observe that this amounts to $\mu=\mu(x,y)$). If we impose the boundary conditions $\mu_1(0,v)=1$ and $\mu_2(0,v)=1/(kv^2-\alpha^2)$, 
 we easily obtain, respectively,  the Jacobi multipliers
$$
\mu_1 = \frac{1}{1+kx^2},\qquad\qquad \mu_2=\frac1{kv^2-\alpha^2}.
$$	
Both Jacobi multipliers lead to the existence of a constant of motion for $\Gamma$ of the form
\begin{equation}\label{Integral1}
I=\frac{\mu_2}{\mu_1}=\frac{1+kx^2}{kv^2-\alpha^2}.
\end{equation}
Note that, as previously remarked, a constant of motion $I$  for (\ref{SODE}) and a Jacobi multiplier $\mu_1$ 
enable us to recover all the Jacobi multipliers for $X$ as $\mu=G(I) \mu_1 $ with $G(I)$ being an appropriately non-vanishing function.

Let us now turn to working out a Lagrangian for oscillator (\ref{NonL1}) by using the method of Jacobi multipliers and $\mu_1$ (observe that $\mu_1$ is also a Jacobi multiplier of (\ref{NonL1}) for $k=0$). Recall that this method states that (\ref{NonL1}) admits a Lagrangian $L_1:{\rm T}\mathbb{R}\rightarrow\mathbb{R}$ satisfying
$$
  \frac{\partial^2 L_1}{\partial v^2}=\mu_1=\frac{1}{1+kx^2}.
$$
This yields
$$
L_1(x,v) = \frac{1}{2} \frac{v^2}{(1+kx^2)}+\phi_1(x)\,v+\phi_2(x)  \,,
$$
for certain functions $\phi_1, \phi_2:\mathbb{R}\rightarrow \mathbb{R}$. We can set the gauge term $\phi_1(x)\,v$ equal to zero while $\phi_2$ is to be determined using (\ref{phidos2}). More specifically, choosing $v_0=0$ in Theorem \ref{Main1}, we obtain that the function $\phi_2 $ is given, up to the addition of a constant, by
$$
\begin{aligned}
\phi_2(x) &=\int_0^x (\mu\,F)(\zeta,0)\,{\rm d}\zeta=\int_0^x\frac{1}{1+k\zeta^2}\,\frac{(-\alpha^2 )\zeta}{1+k\zeta^2}{\rm d}\zeta=\left[\frac{\alpha^2}{2k}\,\frac{1}{1+k\zeta^2}\right]_0^x=-\frac{\alpha^2\,x^2}{2(1+kx^2)}\,. 
\end{aligned}
$$
Therefore the Lagrangian for (\ref{NonL1}) is given, up to addition of a gauge term,  by 
\begin{equation*}
L_1(x,v)= \frac{1}{2} \frac{v^2-\alpha^2 x^2}{1+kx^2} \,, 
\end{equation*}
and the corresponding momentum and Hamiltonian read
\begin{equation*}
p=\frac{\partial L_1}{\partial x}=\frac{v}{1+kx^2}\qquad \Rightarrow \qquad H_1(x,p) = \frac{1}{2} (1+kx^2)p^2+\frac{1}2\frac{\alpha^2\,x^2}{1+kx^2}  \,.  
\end{equation*}
We note that the function $L_1$ coincides with the Lagrangian obtained in \cite{ML74} by direct approach. It is remarkable that $L_1$ is a standard mechanical Lagrangian: it is given by a kinetic term quadratic in the velocities minus a potential term. Moreover,
we can also prove that $L_1$ is, up to an irrelevant additive constant, the Lagrangian function $L$ for $\Gamma$ obtained by using Proposition \ref{LagInt} and the constant of motion $I=\mu_1/(2k\mu_2)$:
$$
L(x,v)=v\int^v\frac{I(x,\zeta)}{\zeta^2}{\rm d}\zeta=v\int^v\frac{k\zeta^2-\alpha^2}{2k(1+kx^2)\zeta^2}{\rm d}\zeta=\frac{kv^2+\alpha^2}{2k(1+kx^2)}=L_1+\frac{\alpha^2}{2k}.
$$

Meanwhile, the second Jacobi multiplier, $\mu_2=(kv^2-\alpha^2)^{-1}$, gives rise to a non-mechanical Lagrangian. Indeed, if we assume $k>0$ and $\partial^2L_2/\partial v^2=\mu_2$, we obtain that, up to irrelevant gauge terms, the corresponding Lagrangian reads
$$
L_2(x,v)=-\frac{v}{\sqrt{k}\alpha}{\rm arcth}\left(\frac{\sqrt{k}v}{\alpha}\right)-\frac{1}{2k}\ln |\alpha^2-kv^2|+\phi_2(x),
$$
where 
$$
\phi_2(x)=\int^x_0(\mu F)(\zeta,0){\rm d}\zeta=\int^x_0\,\frac{\zeta {\rm d}\zeta}{1+k\zeta^2}=\frac{\ln (1+kx^2)}{2k}.
$$
Hence,
$$
L_2(x,v)=-\frac{v}{\sqrt{k}\alpha}{\rm arcth}\left(\frac{\sqrt{k}v}{\alpha}\right)+\frac1{2k}{\ln \frac{1+kx^2}{|\alpha^2-kv^2|}}.
$$
In consequence, 
$$
p=-\frac{{\rm arcth}(\sqrt{k}v/\alpha)}{\sqrt{k}\,\alpha}\Rightarrow
H_2(x,p)=\frac{1}{2k}\ln\left(\frac{\alpha^2\left[1-{\rm th}(\sqrt{k}\,p\,\alpha)^2\right]}{1+kx^2}\right).
$$
The case $k<0$ leads to a similar result.
\subsection{Second nonlinear oscillator }

We now apply Jacobi multipliers to nonlinear oscillators (\ref{NonL2}). Proceeding as before, we consider such systems as  first-order systems by adding a new variable $v\equiv {\rm d}x/{\rm d}t$ to obtain
\begin{equation}\label{AsoNonL2}
\left\{
\begin{aligned}
\frac{{\rm d}x}{{\rm d}t}&=v\,,\\
\frac{{\rm d}v}{{\rm d}t}&=-\frac{kxv^2}{1+kx^2}-\frac{\alpha^2x}{(1+kx^2)^3}\,.
\end{aligned}\right.
\end{equation}
We drop the case $k=0$ as it leads to the  standard harmonic oscillator which was analysed in previous section. So, we now focus upon the case $k\neq 0$.
The multiplier equation (\ref{multcond1}) for the vector field associated to the above system reads
$$
v\frac{\partial \mu}{\partial x}-\left(\frac{kxv^2}{1+kx^2}+\frac{\alpha^2x}{(1+kx^2)^3}\right)\frac{\partial \mu}{\partial v}-\frac{2\mu kxv}{1+kx^2}=0\,,
$$
whose {\it characteristic system} is
$$
\frac{{\rm d}x}{v}=-\frac{(1+kx^2)^3\, {\rm d}v}{kxv^2(1+kx^2)^2+\alpha^2x}=\frac{(1+kx^2)\,{\rm d}\mu}{2 kxv\mu}\,.
$$
The equality between the first and the last term shows that
$$
\frac{{\rm d}x}{1+kx^2}=\frac{{\rm d}\mu}{2kx\mu}\Rightarrow \frac{\mu}{1+kx^2}=\Upsilon_1
$$
for a real constant $\Upsilon_1$. Meanwhile, the integration of the equation
$$
\frac{{\rm d}x}{v}=-\frac{(1+kx^2)^3\,  {\rm d}v}{kxv^2(1+kx^2)^2+\alpha^2x}
$$
goes as follows. Rewrite the above equation as
$$
-\frac{kv^2(1+kx^2)^2+\alpha^2}{(1+kx^2)^3}=\frac{v\,{\rm d}v}{x\, {\rm d}x}=\frac{{\rm d}v^2}{{\rm d}x^2}\,.
$$
The local change of variables $w\equiv v^2$, $z\equiv x^2$ transforms the above equation into 
$$
\frac{{\rm d}w}{{\rm d}z}=-\frac{k}{1+kz}w-\frac{\alpha^2}{(1+kz)^3}\,,
$$
which finally gives
$$
\frac{k(1+kx^2)^2v^2-\alpha^2}{k(1+kx^2)}=\Upsilon_2\,,
$$
for a certain real constant $\Upsilon_2$. Resuming, we obtain the characteristic curves
$$
\frac{\mu}{1+kx^2}=\Upsilon_1,\qquad \frac{k(1+kx^2)^2v^2-\alpha^2}{k(1+kx^2)}=\Upsilon_2\,.
$$
Imposing, for instance, $\mu(0,v)=1$ or $\mu(0,v)=(kv^2-\alpha^2)/k$, we obtain
$$
\mu_1=1+kx^2,\qquad \mu_2={k(1+kx^2)^2v^2-\alpha^2}\,,
$$
which enable us to define a constant of motion
\begin{equation}\label{Integral2}
I=\frac{\mu_2}{\mu_1}=\frac{k(1+kx^2)^2v^2-\alpha^2}{1+kx^2}\,.
\end{equation}
One of the reasons of the interest of our results, in particular of the first-integrals (\ref{Integral1}) and (\ref{Integral2}), is that they provide a new geometric approach to results provided previously in  \cite{BGS11}. Additionally, along with Proposition \ref{LagInt}, it enables us to obtain new Lagrangian descriptions of the oscillators under study.

Let us work out a Lagrangian for (\ref{NonL2}) via $\mu_1$. From equation
$$
\frac{\partial^2 L_1}{\partial v^2}=\mu_1=1+kx^2,
$$
we obtain
$$
L_1(x,v) = \frac{1}{2} (1+kx^2)\,v^2 + \bar \phi_1(x)v+\bar \phi_2(x),
$$
for certain functions $\bar \phi_1$ and $\bar \phi_2$. The gauge term $\bar{\phi}_1(x)v$ can be set to zero 
and  $\bar \phi_2(x) $ is   determined by (\ref{phidos2}) where we  choose $v_0=0$, i.e.
$$
\begin{aligned}
\bar \phi_2(x) &=\int_0^x (\mu\,F)(\zeta,0)\,{\rm d}\zeta=\int_0^x\frac{(-\alpha^2\,\zeta )}{(1+k\zeta^2)^2}\,{\rm d}\zeta=\left[\frac{\alpha^2}{2k(1+kx^2)}\right]_0^x=-\frac{\alpha^2\,x^2}{2(1+kx^2)}\,.
\end{aligned}
$$
Then,  the Lagrangian for (\ref{NonL2}) is given, up to addition of a gauge term,  by 
\begin{equation*}
  L_1(x,v)=\frac{1}{2}(1+kx^2)\,v^2 - \frac{1}{2}\frac{\alpha^2\,x^2}{1+kx^2} \,, 
\end{equation*}
with  corresponding Hamiltonian
\begin{equation*}
  H_1(x,p) = \frac{1}{2}\frac{p^2}{1+kx^2}+\frac{1}{2}\frac{\alpha^2\,x^2}{1+kx^2}  \,. 
\end{equation*}

Meanwhile, if we make use of the second Jacobi multiplier, $\mu_2=k(1+kx^2)^2v^2-\alpha^2$, we obtain a non-standard Lagrangian, because
$$
\frac{\partial^2 L_2}{\partial v^2}=\mu_2\Longrightarrow L_2(x,v)=\frac{k}{12}v^4(1+kx^2)^2-\frac{v^2\alpha^2}{2}+\bar{\phi}_2(x),
$$
with
$$
\bar{\phi}_2(x)=\int_0^x (\mu\,F)(\zeta,0)\,{\rm d}\zeta=\int^x_0\frac{\alpha^4\zeta}{(1+k\zeta^2)^3}{\rm d}\zeta=\frac{\alpha x^2(2+k x^2)}{4(1+kx^2)^2}.
$$
Hence, up to an irrelevant gauge term, we obtain that
$$
L_2(x,v)=\frac{k}{12}v^4(1+kx^2)^2-\frac{v^2\alpha^2}{2}+\frac{\alpha x^2(2+k x^2)}{4(1+kx^2)^2}.
$$
Consequently, 
$$
p=\frac k3(1+kx^2)^2v^3-\alpha^2v\Rightarrow v=\pm\frac{\sqrt{p+\alpha^2}}{\sqrt{k+2k^2x^2+k^3x^4}}
$$
and
$$ 
H_2(x,p)=\frac{p^2+4\alpha^2+\alpha(-kx^2(2+kx^2)+3\alpha^3-4\alpha\sqrt{k(1+kx^2)^2(p+\alpha^2)}}{4k(1+kx^2)^2}.
$$
\section{Jet bundle formulation of  symmetries of differential equations}

Systems of differential equations and their symmetries admit an alternative geometric approach: instead of considering them as vector fields, we describe them as geometric structures in jet bundles. 
We now recall the basic ingredients of this formulation  before passing to study non-local symmetries of differential equations.
 
Let $(E,\mathbb{R},\pi)$ be a fiber bundle with total space $E\equiv \mathbb{R}\times N$, base $\mathbb{R}$ and submersion $\pi:(t,x)\in E\mapsto t \in \mathbb{R}$. Given local coordinates $\{x^j\}_{j=1,\ldots,n}$ on $N$ and $t$ on $\mathbb{R}$, we can naturally define a coordinate system $\{t,x^j_{0)}\equiv x^j\}_{j=1,\ldots,n}$ on $E$. Given a section $\sigma:\mathbb{R}\rightarrow E$ of $(E,\mathbb{R},\pi)$, we write $j^k_{t_0}\sigma=(t_0,x_{0)},x_{1)},\ldots,x_{k)})$ for the $k$-order {\it jet prolongation} of $\sigma$ at $t_0$, i.e. the equivalence class of sections $\gamma:t\in\mathbb{R}\mapsto (t,\gamma^1(t),\ldots,\gamma^n(t))\in E$ such that
$$
\gamma^j(t_0)=x^j_{0)},\,\,\,\quad \frac{{\rm d}^i\gamma^j}{{\rm d}t^i}(t_0)=x^j_{i)},\,\,\, \quad\qquad i=1,\ldots,k,\quad\,\,\, j=1,\ldots,n\,.
$$
We denote by $J^k\pi$, for $k\geq 0$, the space of $k$-order jet prolongations ($k$-jets) of the fiber bundle $(E,\mathbb{R},\pi)$ and we define $J^0\pi\equiv E$. Alternatively, we write $J^k(\mathbb{R},E)$ for $J^k\pi$ when $\pi$ is understood from the knowledge of $E$ and $\mathbb{R}$. The space $J^k\pi$ is a finite-dimensional manifold with local coordinates $\{t,x^j_{i)}\}_{\stackrel{i=0,\ldots,k}{j=1,\ldots,n}}$ of the form
$$
t(j^k_{t_0}\gamma)=t_0,\qquad
x^j_{i)}(j^k_{t_0}\gamma)=\frac{{\rm d}^i\gamma^j}{{\rm d}t^i}(t_0),\qquad\qquad i=0,\ldots,k,\quad\,\,\, j=1,\ldots,n\,,
$$
with ${\rm d}^0\gamma^j/{\rm d}t^0(t_0)\equiv \gamma^j(t_0)$.
It is well known that $(J^k\pi,\mathbb{R},\pi_k:J^k\pi\rightarrow \mathbb{R})$ is a fiber bundle: the {\it $k$-order  jet bundle} associated with $(E,\mathbb{R},\pi)$. The sections of the $k$-order jet bundle being the prolongation of a section of $E$ are called ($k$-order) {\it holonomic sections}\cite{DS89,KS08}.

Consider the $C^\infty(J^k\pi)$-module of 1-forms $\theta\in \Lambda^1(J^k\pi)$ satisfying that $(j^k\sigma)^*\theta=0$ for every section $\sigma:\mathbb{R}\rightarrow E$. The elements of this module are called {\it contact forms} on $J^k\pi$. It can be proved that this module is a locally free-module generated by the contact forms  $\theta^j_{i)}\equiv{\rm d}x^j_{i)}-x^{j}_{i+1)}\,{\rm d}t$, with $j=1,\ldots,n$ and  $i=0,\ldots, k-1$.
We can endow $J^k\pi$ with the distribution $\mathcal{C}^k$ spanned by vector fields annihilating all contact forms on $J^k\pi$,  the referred to as {\it contact or Cartan distribution} of $J^k\pi$. In particular, tangent vectors to graphs of $k$-order jet prolongations belong to $\mathcal{C}^k$. More generally, the Cartan distribution of $J^k\pi$ is spanned by 
\begin{equation}\label{gen}
D_{k)}=\frac{\partial}{\partial t}+\sum_{j=1}^n\sum_{i=0}^{k-1} x^{j}_{i+1)}\frac{\partial}{\partial x^j_{i)}},\quad\,\,\, D_{j}=\frac{\partial}{\partial x_{k)}^j},\,\,\, \qquad j=1,\ldots,n.
\end{equation}
Note that $\mathcal{C}^{k}$ is not involutive for finite $k$ and have dimension $n+1$.

We call {\it Lie symmetries}\cite{Ca07} of $\mathcal{C}^{k}$  the infinitesimal symmetries of $\mathcal{C}^k$, i.e. the vector fields $Y$ on $J^k\pi$ satisfying  that $[Y,X]$ takes values in $\mathcal{C}^k$ for every vector field $X$ taking values in $\mathcal{C}^k$. In other words, Lie symmetries of $\mathcal{C}^k$ are those vector fields whose flows give rise to transformations mapping $k$-order holonomic sections into $k$-order holonomic sections. 

Given a vector field $X$ on $E$, its {\it prolongation} to $J^k\pi$ is the unique Lie symmetry, $X^{(k)}$, of $\mathcal{C}^k$ whose holonomic integral curves are
 the prolongations to $J^k\pi$ of integral curves of $X$. Equivalently, $X^{(k)}$ is the unique Lie symmetry of $\mathcal{C}^k$ projecting onto $X$ under $\pi_{k,0}:j^k_t\sigma\in J^k\pi\mapsto \sigma(t)\in E$.
{\it Lie point symmetries} of $\mathcal{C}^k$ are Lie symmetries of $\mathcal{C}^k$ that are prolongations  to $J^k\pi$ of vector fields on $E$. 
The Lie--B\"acklund theorem states that not all Lie symmetries of $\mathcal{C}^k$ are Lie point symmetries. Non-Lie point symmetries, the referred to as {\it contact Lie symmetries}, can always 
be considered as liftings of a uniquely defined  Lie symmetry on $J^1\pi$ (see \cite{Ca07}).

In the above framework, a $k$-order system of differential equations is a closed embedded submanifold $\mathcal{E}\subset J^k\pi$ and its particular solutions are sections of $\pi$
 whose $k$-order prolongations belong to $\mathcal{E}$. We say that a system of $k$-order differential equations is in normal form when the natural projection
  $\pi_{k,k-1}:j^k\sigma\in\mathcal{E}\mapsto j^{k-1}\sigma\in J^{k-1}\pi$ is a submersion, e.g. $\mathcal{E}=\{j^2_tx\in J^2(\mathbb{R},\mathbb{R}^3):x^1_{2)}-x^2_{2)}=0\}$. 
  We say that $\mathcal{E}$ is in normal form and not underdetermined when $\pi_{k,k-1}:j^k\sigma\in\mathcal{E}\mapsto j^{k-1}\sigma\in J^{k-1}\pi$ is a diffeomorphism, 
  e.g. $\mathcal{E}=\{j^2_tx\in J^2(\mathbb{R},\mathbb{R}^3)\mid x^j_{2)}=F^j(t,x,x_{1)}),j=1,2\}$ for arbitrary functions $F^1,F^2:J^1(\mathbb{R},\mathbb{R}^3)\rightarrow \mathbb{R}$. 
  When $\mathcal{E}\subset J^k\pi$ is in normal form and not underdetermined, there exists a vectorial mapping $\Delta:J^k\pi\rightarrow \mathbb{R}^n$ allowing us to write that $\mathcal{E}=\Delta^{-1}(0)$. 

A Lie symmetry of $\mathcal{C}^k$ that is tangent to $\mathcal{E}$ is called a {\it classical symmetry of $\mathcal{E}$} \cite{Ca07}. On the other hand, the term classical symmetry of $\mathcal{E}$ has also being employed \cite{Vi89} to refer to a vector field on $E$ giving rise to a uniparametric group of transformations mapping particular solutions to $\mathcal{E}$ to particular solutions to $\mathcal{E}$. In this work we will mainly use the first definition. Nevertheless, if a classical symmetry $Y$ is a Lie point symmetry, then we also call classical symmetry the unique vector field on $E$ whose prolongation to $J^k\pi$ is $Y$. More specifically, we say that $Y$ is a {\it classical point symmetry} of $\mathcal{E}$.

The projections $\pi_{h,k}:J^h\pi\rightarrow J^k\pi$, with $h>k$, enable us to define the bundle of infinite jets $(J^{\infty}\pi,\mathbb{R},\pi_\infty:J^{\infty}\pi\rightarrow \mathbb{R})$ as the inverse limit of the projections
$$
\mathbb{R}\leftarrow E\leftarrow J^1\pi\leftarrow J^2\pi\leftarrow\ldots
$$
The commutative ring  of differentiable functions over $J^\infty\pi$ is defined by $\mathcal{F}(\pi)\equiv\bigcup_{l=0}^\infty C^{\infty}(J^l\pi)$. Note that each element of $\mathcal{F}(\pi)$ depends on a finite subset of variables of $\{t,x^j_{i)}\}_{\stackrel{i\in \mathbb{N}\cup \{0\}}{j=1,\ldots,n}}$. When there exists a natural injection between two manifolds, e.g.  $J^k\pi\hookrightarrow J^{k'}\pi$ for $k'>k$, we may consider each function on the first manifold as a function on the second, e.g. an element of $C^\infty(J^k\pi)$ as an element of $C^\infty(J^{k'}\pi)$, so as to simplify the notation.

Given a section $\sigma:\mathbb{R}\rightarrow J^k\pi$, we call {\it infinite prolongation} of $\sigma$ the section $j^\infty\sigma:t\in\mathbb{R}\mapsto j^\infty_t\sigma\in J^\infty\pi$ given in coordinates by
$$
j_t^\infty\sigma\equiv\left(t,\sigma(t),\frac{{\rm d}\sigma}{{\rm d}t}(t),\frac{{\rm d}^2\sigma}{{\rm d}t^2}(t),\ldots,\right).
$$

Similarly to  finite-dimensional manifolds, vector fields on $J^\infty\pi$ are defined as derivations of the commutative ring $\mathcal{F}(\pi)$ and the $\mathcal{F}(\pi)$-module of vector fields on $J^\infty\pi$ becomes a  Lie algebra with respect to the commutator of derivations. The 
tangent vectors to infinite prolongations of sections of $E$ span a distribution on $J^\infty \pi$ spanned  by the derivation on $\mathcal{F}(\pi)$ given by 
$$
  D=\frac{\partial}{\partial t}+\sum_{j=1}^n\sum_{i=0}^{\infty} x^{j}_{i+1)}\frac{\partial}{\partial x^j_{i)}}.
$$
Although $D$ depends on an infinite number of variables, it induces a well-defined derivation on $\mathcal{F}(\pi)$ due to the fact that every function in $\mathcal{F}(\pi)$ only depends on a finite number of variables. The distribution $\mathcal{C}$ spanned by $D$ is the referred to as {\it contact or Cartan distribution} of $J^\infty\pi$, which is one-dimensional and therefore involutive. We cannot  ensure that $D$ is integrable as the Frobenius Theorem does not apply to distributions on infinite-dimensional manifolds. 

We call Lie symmetries of $\mathcal{C}$ the infinitesimal symmetries of $\mathcal{C}$. As these Lie symmetries of $\mathcal{C}$ are defined on an infinite-dimensional manifold, we cannot ensure that they are associated to uni-parametric groups of diffeomorphisms on $J^\infty\pi$.

It is interesting to note that many structures on the jet spaces $J^k\pi$ become simpler when passing to $J^\infty\pi$, e.g. the Cartan distribution becomes one-dimensional and involutive. Moreover, $J^\infty\pi$ is geometrically richer  than finite-order jet bundles. For instance, Lie symmetries of $\mathcal{C}$ need not be lifts  neither of vector fields on $E$ nor of vector fields on $J^1\pi$ due to the fact that the Lie--B\"acklund theorem does not apply to such Lie symmetries \cite{Ca07}. 

Given a $k$-order system of differential equations $\mathcal{E}\subset J^k\pi$, the {\it $l$-prolongation of $\mathcal{E}$} is the set of points $\mathcal{E}^{(l)}\equiv\{j^{k+l}_t\sigma\,|\,j^k\sigma\subset \mathcal{E}\}\subset J^{k+l}\pi$ (see \cite{Ca07,Vi89} for details). Further, we call {\it infinite prolongation of $\mathcal{E}$} the set $\mathcal{E}^{\infty}\equiv\{j^{\infty}_t\sigma\,|\,j^k\sigma\subset \mathcal{E}\}\subset J^{\infty}\pi$. If $\mathcal{E}$ is in normal form and not underdetermined, then $\mathcal{E}=\Delta^{-1}(0)$ for  a certain mapping $\Delta:j^k_tx\in J^k\pi\mapsto (x^1_{k)}-F^1(j_t^{k-1}x),\ldots,x^n_{k)}-F^n(j_t^{k-1}x))\in  \mathbb{R}^n$ and functions $F^j:J^{k-1}\pi\rightarrow \mathbb{R}$. In this case, $\mathcal{E}^\infty$ is a finite-dimensional manifold locally determined by the infinite set of conditions
\begin{equation}\label{conCon}
\Delta=0,\qquad D^s\Delta=\stackrel{s-{\rm times}}{\overbrace{D(D(\ldots D(D}}\Delta)\ldots))=0\,,
\end{equation}
with $s=1,2,\ldots$ The above conditions determine all derivatives of particular solutions of $\mathcal{E}$ out of the value of the first $(k-1)$-derivatives.  
Hence, a local set of coordinates for $\mathcal{E}$ can be considered in a natural way as  elements of $\mathcal{F}(\pi)$ giving rise to a coordinate system on $\mathcal{E}^\infty$, which becomes a finite-dimensional manifold. In view of (\ref{conCon}), the restriction of $D$ to $\mathcal{E}^\infty$ is tangent to $\mathcal{E}^\infty$. This allows us to endow $\mathcal{E}^\infty$ with a one-dimensional distribution $\mathcal{C}|_{\mathcal{E}^\infty}$.  The pair given by $\mathcal{E}^\infty$ and $\mathcal{C}|_{\mathcal{E}^\infty}$ becomes what is called a one-dimensional {\it diffiety}. We define $\mathcal{F}(\mathcal{E}^\infty)$ to be the restriction to $\mathcal{E}^\infty$ of functions of $\mathcal{F}(\pi)$. The Lie symmetries of $\mathcal{C}$ that are tangent to $\mathcal{E}^\infty$ are called {\it higher symmetries} of $\mathcal{E}$.

\begin{definition} 
Let $\mathcal{E}$ be a $k$-order system of ODEs, we say that a bundle $(\mathcal{E}^c,\mathcal{E}^\infty,\kappa:\mathcal{E}^c\rightarrow\mathcal{E}^\infty)$ is a {\it covering} for $\mathcal{E}$ if the bundle ${\mathcal{E}}^c$ can be endowed with a one-dimensional distribution
$$
\mathcal{C}^c=\{\mathcal{C}^c_p\}_{p\in \widetilde{\mathcal{E}}}
$$
in such a way that $(\kappa_*)_p:{\mathcal{C}^c_p}\rightarrow \mathcal{C}_{\kappa(p)}$ is a linear isomorphism for each $p\in{\mathcal{E}^c}$.
\end{definition}

\begin{definition} Let $\mathcal{E}$ be a $k$-order system of ODEs and let ${\mathcal{E}^c}$ be a covering for $\mathcal{E}^\infty$. We call {\it non-local symmetry} for $\mathcal{E}$ an infinitesimal symmetry of the distribution ${\mathcal{C}^c}$.
\end{definition}

We call dimension of the covering, $\dim(\kappa)$, the dimension of the fiber of the bundle $\kappa$.
Given a covering $\kappa$, integral manifolds $\mathcal{S}$ of ${\mathcal{C}^c}$ project under $\kappa$ onto integral manifolds of $\mathcal{C}|_{\mathcal{E}^\infty}$, i.e. onto prolongations of
particular solutions of $\mathcal{E}$. It follows that, in this picture, symmetries
of ${\mathcal{C}^c}$ shuffle integral manifolds of ${\mathcal{C}^c}$ and, projecting under $\kappa$,  we can obtain particular solutions to $\mathcal{E}$.

\section{Extended formalism and non-local symmetries}

Recently Gandarias and coworkers have studied a method for obtaining non-local symmetries for certain particular systems of ODEs \cite{BGS11, Ga09, GB11}. The main idea is to add a new additional equation to the original one in such a way that the new higher-dimensional system can be endowed with a classical symmetry. 
In this section, we study this technique from a geometric perspective. 
The starting point is that this procedure must be considered, in geometric terms, as a very particular case of an extended formalism where the covering has a total space given by a one-dimensional diffiety. We also mention the relation of this approach with some results previously obtained by   Krasil'shchi, Vinogradov and coworkers \cite{V99,Ca07,KKV04}.

We begin with a system $\mathcal{E}$ of $k$-order ODEs
 given by $\mathcal{E}=\Delta^{-1}(0)\subset J^k{\tau_{n+1}}$ for a mapping $\Delta:J^k\tau_{n+1}\rightarrow \mathbb{R}^p$ with $(\mathbb{R}^{n+1},\mathbb{R},{\tau}_{n+1}:(t,x)\in\mathbb{R}\times\mathbb{R}^{n}\mapsto t\in \mathbb{R})$ and coordinates
\begin{equation}\label{original}
  \Delta^i\left(t,x,\frac{{\rm d}x}{{\rm d}t},\ldots,\frac{{\rm d}^kx}{{\rm d}t^k}\right)=0,\quad i=1,\ldots,p,\quad x\in\mathbb{R}^n
\end{equation}
and we assume that $\mathcal{E}$ is underdetermined and in normal form, i.e. $n=p$ and $\pi_{k,k-1}:\mathcal{E}\subset J^k\pi\rightarrow J^{k-1}\pi$ is a diffeomorphism.
A classical point symmetry of $\mathcal{E}$, represented by an  $\epsilon$-parametric group of transformations given infinitesimally by
$$
\left\{\begin{aligned}
\bar t& = t+\epsilon\,\xi(t,x)  \,,\\
\bar{x}^j& = x^j+\epsilon\,\eta^j(t,x)  \,,\\
\end{aligned}\right.\qquad j=1,\ldots,n,
$$
preserves the set of solutions of the equation, that is,  transforms particular solutions of  (\ref{original}) into particular solutions of the same  equation.  Unfortunately, many systems of  differential equations do not possess classical point symmetries. In that case, the method of non-local symmetries (applied to some particular cases in \cite{BGS11, Ga09, GB11} and related to some questions discussed in \cite{V99,Ca07,KKV04}) can be of a great usefulness.

A {classical point symmetry}  for the system (\ref{original}) is a vector field $Y$ on $J^k{\tau}_{n+1}$ such that (i) is the lift of a vector field $Y_0$ on $J^0{\tau}_{n+1}\simeq \mathbb{R}^{n+1}$, and (ii) is tangent to the submanifold $\mathcal{E}$.  
In coordinates if $Y_0$ takes the form 
$$
 Y_0(t,x)=\xi(t,x)\frac{\partial}{\partial t}+\sum_{j=1}^n\eta^j(t,x)\frac{\partial}{\partial x^j},
$$
then $Y$ is given by 
\begin{equation}\label{prol}
 Y=\xi\frac{\partial}{\partial t}+\sum_{j=1}^n\left(\eta^j\frac{\partial}{\partial x^j} + \sum_{i=1}^k\varphi^{(i)}_{\xi,\eta}\frac{\partial}{\partial x^j_{i)}}\right),
\end{equation}
where the functions 
$\varphi^{(i)}_{\xi,\eta}:J^k\tau_{n+1}\subset J^\infty \tau_{n+1}\rightarrow\mathbb{R}$ 
can be obtained from the functions $\xi$ and $\eta^j$, with $j=1,\ldots,n$ (for a detailed explanation see\cite{St89}).

Suppose that the system (\ref{original}) is given.  Let us construct a new system containing (\ref{original}) as a particular part. Consider the new jet bundles associated to the bundle $(\mathbb{R}^{n+2},\mathbb{R},\tau_{n+2}:(t,\bar x)\in\mathbb{R}^{n+2}\mapsto t\in \mathbb{R})$, with $\bar x\equiv (x,w)\in \mathbb{R}^{n+1}$. For a certain fixed function $H:J^1\tau_{n+1}\rightarrow \mathbb{R}$,
we can define a new system
$\widetilde{\mathcal{E}}\subset J^k\tau_{n+2}$ as follows 
\begin{equation}\label{cov}
 \Delta\left(t,x,\frac{{\rm d}x}{{\rm d}t},\ldots,\frac{{\rm d}^{k}x}{{\rm d}t^k}\right)=0,\quad
\frac{{\rm d} w}{{\rm d}t}=H, \quad\frac{{\rm d}^2 w}{{\rm d}t^2}=D_{k)}H,\quad\ldots,\quad\frac{{\rm d}^k w}{{\rm d}t^k}=D^{k-1}_{k)}H,
\end{equation}
containing as a particular part the initial system (\ref{original}) and where $D_{k)}$ is one of the generators of the Cartan distribution of $J^k\tau_{n+2}$ given in (\ref{gen}).  Observe that although $\mathcal{E}$ was in normal form, the new system (\ref{cov}) is not in normal form as well: the derivatives ${\rm d}^kw/{\rm d}t^k$ can be expressed in terms of the lower derivatives, but such lower derivatives must hold several additional conditions. More geometrically, we cannot consider (\ref{cov}) straightforwardly as a submanifold $\widetilde{\mathcal{E}}\subset J^k\tau_{n+2}$ in such a way that $\tau_{k,k-1}:\widetilde{\mathcal{E}}\subset J^{k}\tau_{n+2}\rightarrow J^{k-1}\tau_{n+2}$ is epijective.

In any case, we can consider the prolongations $\mathcal{E}^\infty$  and $\widetilde{\mathcal{E}}^\infty$ of $\mathcal{E}$ and $\widetilde {\mathcal{E}}$, respectively. Since the system $\mathcal{E}$ is in normal form and in view of the definition of $\widetilde{\mathcal{E}}$, we see that the values of each derivative ${\rm d}^{\bar p}x/{\rm d}t^{\bar p}$, with $\bar{p}>k$, and ${\rm d}^{\bar p}w/{\rm d}t^{\bar p}$, with $\bar p>1$, of each particular solution of $\mathcal{E}$ and $\widetilde{\mathcal{E}}$ can be  determined from the previous derivatives. Thus, $\mathcal{E}^\infty$ and $\widetilde{\mathcal{E}}^\infty$ are finite-dimensional manifolds and
$$
 \dim\widetilde{\mathcal{E}}^\infty = \dim\widetilde{\mathcal{E}} \,,\qquad 
 \dim\mathcal{E}^\infty = \dim\mathcal{E}.
$$
Hence, a local coordinate system on $\mathcal{E}$ or $\widetilde{\mathcal{E}}$ induces a local coordinate system on their infinite prolongations. This means that expressions in coordinates on $\mathcal{E}$ and $\widetilde{\mathcal{E}}$ can be understood as expressions on $\mathcal{E}^\infty$ or $\widetilde{\mathcal{E}}^\infty$ indistinctly. This property is important: it allows us to identify $\mathcal{E}^\infty$ with $\mathcal{E}$ and $\widetilde{\mathcal{E}}^\infty$ with $\widetilde{\mathcal{E}}$. Hence, calculations in infinite-dimensional jet bundles are as difficult as for finite-dimensional jet bundles and the whole procedure is properly defined in a rigorous more powerful geometrical manner. For instance, $D$, which has no sense on $J^k\tau_{n+1}$, can be however correctly considered when restricted to $\mathcal{E}^\infty$.  

Since $D$ is tangent to $\mathcal{E}^\infty$, we can define the restriction $D|_{\mathcal{E}^\infty}$ of this operator to $\mathcal{E}^\infty$. If $\widetilde{D}$ is the analogue of $D$ on $J^\infty\tau_{n+2}$, this operator is also tangent to $\widetilde{\mathcal{E}}^\infty$ and we can also define its restriction, $\widetilde D|_{\widetilde{\mathcal{E}}^\infty}$, to $\widetilde{\mathcal{E}}^\infty$. The vector field $\widetilde{D}$ induces a one-dimensional distribution $\widetilde{\mathcal{C}}$ on $J^\infty\tau_{n+2}$ and
 $\widetilde D|_{\widetilde{\mathcal{E}}^\infty}$ spans a one-dimensional distribution $\widetilde{\mathcal{C}}|_{\widetilde{\mathcal{E}}^\infty}$, turning the pair $(\widetilde{\mathcal{E}}^\infty,\widetilde{\mathcal{C}}|_{\widetilde{\mathcal{E}}^\infty})$ into what is called a diffiety of dimension one: the dimension one refers to the fact that we have defined a one-dimensional distribution $\widetilde{\mathcal{C}}|_{\widetilde{\mathcal{E}}^\infty}$ on $\widetilde{\mathcal{E}}^\infty$. Moreover, we have the following property.
 
\begin{proposition} 
Let $\widetilde{D}$ be the vector field on $J^\infty\tau_{n+2}$ given by
\begin{equation}\label{restrc}
\widetilde{D} = D+\sum_{i=0}^\infty w_{i+1)}\frac{\partial}{\partial w_{i)}}, \qquad w_0\equiv w.
\end{equation}
Then, we have locally 
\begin{equation}\label{restrc2}
 \widetilde D|_{\widetilde{\mathcal{E}}^\infty}= D|_{\mathcal{E}^\infty}+H\frac{\partial}{\partial w}  \,=\sum_{j=1}^n\left[\sum_{p=0}^{k-2}x^j_{p+1)}\frac{\partial}{\partial x^j_{p)}}+F^j(j^{k-1}_tx)\frac{\partial}{\partial x^j_{k-1)}}\right]+H(j^1_tx)\frac{\partial}{\partial w},
\end{equation}
for certain functions $F^1,\ldots,F^n:\widetilde{\mathcal{E}}^\infty\rightarrow \mathbb{R}$. 
\end{proposition}
\begin{proof} Expression (\ref{restrc}) is trivially the generator of the Cartan distribution of $J^\infty\tau_{n+2}$. 
As we assume $\mathcal{E}$ to be in normal form and not underdetermined, the higher-order derivatives of the particular solutions to $\mathcal{E}$, namely ${\rm d}^kx^j/{\rm d}t^k$, can be locally written as a function of the previous derivatives. So, ${\rm d}^kx^j/{\rm d}t^k=F^j(j^{k-1}_tx)$ for $j=1,\ldots,n$ and certain functions $F^j: J^{k-1}\tau_{n+1}\rightarrow \mathbb{R}$ which is understood in the natural way as a function on $\widetilde{\mathcal{E}}^\infty\subset J^\infty\tau_{n+2}$. 
Using this, recalling that $\{t,x_{i)},w\}_{i=0,\ldots,k-1}$ forms a coordinate system for $\widetilde{\mathcal{E}}^\infty$ and restricting $\widetilde{D}$ from $J^\infty\tau_{n+2}$ to $\widetilde{\mathcal{E}}^\infty$, we obtain that the expression (\ref{restrc2}) follows from (\ref{restrc}).
\end{proof}

The natural projection $\Pi:(t,x,w)\in\mathbb{R}^{n+2}\mapsto (t,x)\in\mathbb{R}^{n+1}$ lifts to
a projection $J^\infty \Pi:J^\infty \tau_{n+2}\rightarrow J^\infty \tau_{n+1}$ satisfying
$$
J^\infty \Pi(j^\infty_t \bar x)=J^\infty \Pi(j^\infty_t (x,w))=j^\infty_t x  \,.
$$
This projection induces a map $J^\infty \Pi|_{\widetilde {\mathcal{E}}^\infty}:\widetilde{\mathcal{E}}^\infty\rightarrow\mathcal{E}^\infty$ obeying that
$$
\left(J^\infty \Pi|_{\widetilde{\mathcal{E}}^\infty}\right)_*(\widetilde D|_{\widetilde{\mathcal{E}}^\infty})=D|_{\mathcal{E}^\infty}.
$$
As a consequence, $J^\infty \Pi$ induces an isomorphism $(J^\infty \Pi|_{\widetilde{\mathcal{E}}^\infty})_{*\xi}:\widetilde{\mathcal{C}}_\xi\rightarrow \mathcal{C}_{J^\infty \Pi(\xi)}$ for every $\xi \in \widetilde{\mathcal{E}}^\infty$ and therefore a covering 
$\kappa_*\equiv (J^\infty \Pi|_{\widetilde{\mathcal{E}}^\infty})_*:\widetilde{\mathcal{C}}|_{\widetilde{\mathcal{E}}^\infty}\rightarrow \mathcal{C}|_{{\mathcal{E}}^\infty}$.
Hence, if the system $\widetilde{\mathcal{E}}$ admits a classical symmetry $Y$, e.g. (\ref{prol}), then $Y$ can be lift to a Lie symmetry $Y^\infty$ of $\widetilde{\mathcal{C}}$, namely a higher symmetry for $\widetilde{\mathcal{E}}$. Both vector fields, $Y$ and $Y^\infty$, are tangent to $\widetilde{\mathcal{E}}$ and $\widetilde{\mathcal{E}}^\infty$, respectively. It is worth noting that due to our assumptions, the coordinate expression in $\mathcal{\widetilde{E}}$ of $Y|_\mathcal{\widetilde{E}}$ and the coordinate expression of $Y^\infty|_{\widetilde{\mathcal{E}}^\infty}$ in $\widetilde{\mathcal{E}}^\infty$ are the same. Moreover, as $Y^\infty$ is a higher symmetry, it leaves invariant $\widetilde{\mathcal{C}}$ and it becomes  a non-local symmetry of $\mathcal{E}^\infty$ when restricted to $\widetilde{\mathcal{E}}^\infty$.

If $\widetilde{\mathcal{E}}$ admits a classical point symmetry $Y$, then we have a one-parametric group of diffeomorphisms given, in infinitesimal form, as
$$
\left\{\begin{aligned}
t^*& = t + \epsilon\,\xi(t,x,w) \,,\\
x^*&= x + \epsilon\,\phi(t,x,w) \,,\\
w^*&= w + \epsilon\,\eta(t,x,w) \,, 
\end{aligned}\right.
$$
transforming solutions to the system $\widetilde{\mathcal{E}}$ into solutions of $\widetilde{\mathcal{E}}$. Hence, the set of transformations
$$
\left\{\begin{aligned}
t^*&= t + \epsilon\,\xi(t,x,w(t))  \,,\\
x^*&= x + \epsilon\,\phi(t,x,w(t))  \,, 
\end{aligned}\right.
$$
 enables us to map particular solutions to $\mathcal{E}$ into solutions of $\mathcal{E}$ by means of the curves $w(t)$ corresponding to particular 
 solutions of $\widetilde{\mathcal{E}}$.

We can summarise the main results proved in this section as follows. 
We have proved we can embed a given system $\mathcal{E}$ into a bigger one whose infinity prolongation has the structure of a diffiety. This structure gives rise to a covering for the  initial system. 
In \cite{Ca07} another covering for the initial system is constructed so as to study it through non-local symmetries.  Nevertheless, that covering cannot be considered neither as a diffiety nor a submanifold of a jet bundle without additional constructions. So, our interpretation is more powerful as we can straightforwardly use classical symmetries of $\widetilde{\mathcal{E}}$ to construct non-local symmetries of $\mathcal{E}$.
Gandarias developed slightly modifications of her method, but all of them can be retrieved as particular cases of the above geometric approach.

\section{Extended formalism  for the nonlinear oscillators}


In this section, we provide a geometric method to construct non-local symmetries for second-order autonomous differential equations. This method is based upon considering our initial second-order differential equations as part of an extended system whose form can be determined out of the initial one. In following subsections we show that this procedure retrieves as a particular case the results given by Gandarias and coworkers. 

\begin{theorem}\label{Th} Every system $\mathcal{E}$ corresponding to 
\begin{equation}\label{ExtSysLem}
\begin{aligned}
\frac{{\rm d}^2x}{{\rm d}t^2}&= F(x,v),
\end{aligned}
\end{equation}
can be extended to a larger system $\widetilde{\mathcal{E}}$ with the additional equation ${\rm d}w/{\rm d}t=H(x,v)$ in such a way that $Y=gX_H$, where $X_H=v\partial_x+F\partial_v+H\partial_w$ is an infinitesimal symmetry of the distribution $\widetilde{\mathcal{C}}|_{\widetilde{\mathcal{E}}^\infty}$ generated by $\partial_t+X_H$ and $g\in \mathcal{F}(\widetilde{\mathcal{E}}^\infty)$ is a first-integral of $X_H,\partial_x,\partial_t$ (as vector fields on $\widetilde{\mathcal{E}}^\infty$). In consequence, $Y$ is a non-local symmetry of (\ref{ExtSysLem}).
\end{theorem}

\begin{proof} We have that $[gX_H,\partial_t+X_H]=-(\partial_tg+X_Hg)X_H$. So, $gX_H$ is an infinitesimal symmetry of $\widetilde{\mathcal{C}}|_{\widetilde{\mathcal{E}}^\infty}$ if and only if $\partial_tg+X_Hg=0$. Under the assumed conditions for $g$, we obtain that $gX_H$ is an infinitesimal symmetry of $\widetilde{\mathcal{C}}|_{\widetilde{\mathcal{E}}^\infty}$. Let us prove that there exists a nonconstant function $g$ satisfying the above conditions.

 Set $H(x,v)\equiv F(x,v)h(v)$ for a certain function $h(v)\neq 0$. Since $\partial g/\partial x=0$, then 
$$
0=X_Hg=F(x,v)\left(\frac{\partial g}{\partial v}+h(v)\frac{\partial g}{\partial w}\right).
$$
If $F\neq 0$ and we require $g$ to be non-constant, then our definition of $H$ ensures the existence of a non-trivial $g$ depending only on $v$ and $w$. If $F=0$, we can choose any $g$ with the required properties of our theorem.
\end{proof}

\begin{corollary} A classical infinitesimal symmetry for the system $\widetilde{\mathcal{E}}$ gives rise to a non-local symmetry of $\mathcal{E}$.
\end{corollary}
\begin{proof} Every classical symmetry for $\widetilde{\mathcal{E}}$ can be extended in view of the Lie--B\"acklund Theorem to a higher symmetry $Y^\infty$ of $\mathcal{\widetilde{E}}$. This higher symmetry is tangent to $\widetilde{\mathcal{E}}^\infty$ and a Lie symmetry of $\widetilde{\mathcal{C}}$, which is, by construction of $\widetilde{\mathcal{E}}^\infty$, tangent to $\widetilde{\mathcal{E}}^\infty$. Hence, $Y^\infty|_{\widetilde{\mathcal{E}}^\infty}$ is a symmetry of $\widetilde{\mathcal{C}}|_{\widetilde{\mathcal{E}}^\infty}$ and it  becomes a non-local symmetry for $\mathcal{E}$. 
\end{proof}

\subsection{First nonlinear oscillator }

Let us review the approach given by Gandarias to study the nonlinear oscillator (\ref{NonL1}). The first-order system (\ref{AsoNonL}) associated to (\ref{NonL1}) is embedded into a new one on ${\rm T}\mathbb{R}\times\mathbb{R}$ of the form
\begin{equation}\label{ExtSys}
\left\{
\begin{aligned}
\frac{{\rm d}x}{{\rm d}t}&= v,\\
\frac{{\rm d}v}{{\rm d}t}&= \frac{(kv^2-\alpha^2)x}{1+kx^2},\\
\frac{{\rm d}w}{{\rm d}t}&=H(x,v),
\end{aligned}\right.
\end{equation}
where $H(x,v)$ is a, undetermined for the moment, function, to be fixed later on.  Let us study the Lie point symmetries of this system.
Particular solutions to system (\ref{ExtSys}) are in a one-to-one correspondence with the integral curves $t\mapsto (t,x(t),v(t),w(t))$ of the vector field 
$$
\bar X_H \equiv \frac{\partial}{\partial t}+X_H\equiv \frac{\partial}{\partial t}+ v\frac{\partial}{\partial x}+ \frac{(kv^2-\alpha^2)x}{1+kx^2}\frac{\partial}{\partial v}+H(x,v)\frac{\partial}{\partial w}  \,. 
$$
Given a vector field $Y=\xi\partial/\partial t+\phi\partial/\partial x+\psi\partial/\partial v+\eta\partial/\partial w$ on $\mathbb{R}^2\times{\rm T}\mathbb{R}$, where we include the time variable $t$, we know that $Y$ determines a Lie point symmetry of this system if $[Y,\partial_t+X_H]=f(\partial_t+X_H)$, where we recall that $t,x,v,w$ are considered as coordinates on $\mathbb{R}^2\times{\rm T}\mathbb{R}$ and $f\in C^\infty(\mathbb{R}^2\times{\rm T}\mathbb{R})$. Equivalently $Y$ is a classical point symmetry of this system 
if its four coefficients satisfy the following equations 
\begin{multline}\label{complsys} 
(Y^{(1)}\Delta^1)_{\Delta=0}=v(kx^2+ 1)\xi_t + vH( kx^2 + 1)\xi_w +xv(kv^2-\alpha^2)\xi_v  
+v^2(kx^2+ 1)\xi_x\\
-x(kv^2-\alpha^2)\phi_v-(kx^2+1)v\phi_x 
- (k x^2+1)\phi_t- H (kx^2+1)\phi_w+ v^2(kx^2+1)\psi  = 0,   \\
\qquad\qquad\qquad\qquad- H_xk\phi x^2 + H (kx^2+1)\eta_w  +\eta_tkx^2-\alpha^2x\eta_v- H_v\psi- H_x\phi  +\eta_x = 0, \\
(Y^{(1)}\Delta^2)_{\Delta=0}=(kx^2+1)(\alpha^2-kv^2)x\xi_t- H( kx^2 +1)x [kv^2-\alpha^2]\xi_w-x^2(kv^2-\alpha^2)^2\xi_v\qquad\qquad\qquad\qquad\\
+ vx[1+kx^2](-kv^2 +\alpha^2)\xi_x
+(k x^2 +1)(kv^2-\alpha^2)x\psi_v  +(kx^2-1) (kv^2-\alpha^2)\phi\\
\qquad\qquad \qquad+(kx^2+ 1)^2v\psi_x - 2kxv(1+kx^2)\psi + H(kx^2+ 1)^2\psi_w+ (k x^2+1)^2\phi_t   = 0\\
(Y^{(1)}\Delta^3)_{\Delta=0}=-H(kx^2+1)\xi_t- H^2(kx^2+ 1)\xi_w- Hx(kv^2-\alpha^2 )\xi_v{\qquad\qquad\qquad\qquad\qquad}\\
- H(kx^2+1)v\xi_x- H_x\xi(kx^2  +1)  + kxv^2\eta_v+(kx^2 + 1)v\eta_x- H_vk\psi x^2\\
\end{multline}
for $\Delta^1=\dot x-v$, $\Delta^2= \dot v-(kv^2-\alpha^2)x/(1+kx^2)$, $\Delta^3=\dot w-H$ and $Y^{(1)}$ being the prolongation to $J^1\tau_4$, with $\tau_4:(t,x,v,w)\in \mathbb{R}^4\mapsto t\in \mathbb{R}$, of the vector field $Y$ on $\mathbb{R}^4$.
We include expressions (\ref{complsys}) to solve several minor typos and mistakes in the previous literature. 
This is a quite difficult system to be solved, which suggests us to assume some kind of simplification. This was done in \cite{BGS11}, whose authors considered as an ansatz a particular form for $\xi,\phi,\psi, \eta$. Now we reconsider this whole approach in a more geometrical and rigorous way.

Equivalently, the differential equation (\ref{NonL1}) can be considered along with the equation ${\rm d}w/{\rm d}t=H(x,v)$. As commented in the latter section, this system can be understood as a submanifold  $\widetilde{\mathcal{E}}$ of $J^2\tau_3$ with $\tau_3:(t,x,w)\in\mathbb{R}^3\mapsto t\in\mathbb{R}$.  
Let us use Theorem \ref{Th} to study the infinitesimal symmetries of $\widetilde{\mathcal{C}}|_{\mathcal{E}^\infty}$. Recall that this amounts to a non-local symmetry for $\mathcal{E}$. 

We can construct a non-local symmetry by assuming $Y=gX_H$ with $H(x,v)=(X_Hv)(x,v)h(v)$ and $g$ being a first-integral of $X_H$ independent of $t$ and $x$, namely, such that   
$$
X_Hg=\frac{(kv^2-\alpha^2)x}{1+kx^2}\left(\,\frac{\partial g}{\partial v}+ h(v)\frac{\partial g}{\partial w}\right)=0,
$$
where we fixed according to Theorem \ref{Th}
$$
H(x,v)=\frac{(kv^2-\alpha^2)x}{1+kx^2}h(v).
$$
By assuming $h(v)=1/v$, we obtain a simple first-integral for $X_H$ of the form $g=e^w/v$. Hence,
$$
Y=e^w\left(\frac{\partial}{\partial x}+H\frac{\partial}{\partial v}+\frac{H}{v}\frac{\partial}{\partial w}\right).
$$
Indeed, observe that $[Y,X_H]=0$. 

As $\{t,x,v,w\}$ can be understood as  coordinates of $\mathbb{R}^2\times {\rm T}\mathbb{R}$ and $\widetilde{\mathcal{E}}$, the vector field $Y$ can also be considered as a vector field on $\mathbb{R}^2\times {\rm T}\mathbb{R}$. In this way, $Y$ is the same Lie symmetry provided in \cite{BGS11}, where it was obtained by the derivation of a particular solution of (\ref{complsys}) using an {\it ad hoc} ansatz for $Y$ and $H$. Meanwhile, we here use a covering to show that Gandarias' and coworkers ansatz corresponds to choose a certain $H$ so that a first-integral for $X_H$ independent of $x,t$ can be obtained. This immediately leads to their same final result.

Note also that we could in principle choose another function $H$ which could potentially lead to different non-local symmetries of $\mathcal{E}$. Nevertheless, the form chosen
 in this work makes computations easier in many cases.

\subsection{Second nonlinear oscillator }

We can  now apply the above method to equations (\ref{NonL2}) to recover the same result provided in \cite{BGS11}. In this new case, the first-order system (\ref{AsoNonL2}) associated to (\ref{NonL2}) is embedded into  one
\begin{equation}\label{ExtSys2}
\left\{
\begin{aligned}
\frac{{\rm d}x}{{\rm d}t}&= v,\\
\frac{{\rm d}v}{{\rm d}t}&= -\frac{kxv^2}{1+kx^2}-\frac{\alpha^2x}{(1+kx^2)^3},\\
\frac{{\rm d}w}{{\rm d}t}&=H(x,v),
\end{aligned}\right.
\end{equation}
on $\mathbb{R}^3$, where $H(x,v)$ is a function to be fixed next. Additionally, we can consider this system as a submanifold $\mathcal{E}\subset J^2\tau_3$ with $\tau_3:(t,x,w)\in\mathbb{R}^3\mapsto t\in\mathbb{R}$. This system describes the integral curves $t\rightarrow (t,x(t),v(t),w(t))$ of the  vector field on $\mathbb{R}^2\times{\rm T}\mathbb{R}\simeq \mathcal{\widetilde{E}}$ of the form
$$
\bar X_H\equiv \frac{\partial}{\partial t}+X_H\equiv v\frac{\partial}{\partial x}-\left(\frac{kxv^2}{1+kx^2}+\frac{\alpha^2x}{(1+kx^2)^3}\right)\frac{\partial}{\partial v}+H\frac{\partial}{\partial w}.
$$
We fix $H$ to be of the previously commented form, i.e.
$$
H(x,v)=-\left(\frac{kxv^2}{1+kx^2}+\frac{\alpha^2x}{(1+kx^2)^3}\right)\frac{1}{v}.
$$
Hence, $X_H$ admits a locally defined first-integral $g$ that does not depend neither on $x$ nor on $t$, namely, such that 
$$
X_Hg=-\left(\frac{kxv^2}{1+kx^2}+\frac{\alpha^2x}{(1+kx^2)^3}\right)\frac{\partial g}{\partial v}+H\frac{\partial g}{\partial w}=0.
$$
This leads to a simple first-integral for $X_H$ of the form $g=e^w/v$. We can now obtain a Lie symmetry of the system by choosing $Y=gX_H$, which reads
$$
Y=e^w\left(\frac{\partial}{\partial x}+H\frac{\partial}{\partial v}+\frac{H}v\frac{\partial}{\partial w}\right),
$$
which is again the same classical symmetry provided in \cite{BGS11} but we here understand it as a symmetry of $\widetilde{\mathcal{C}}$ on $\widetilde{\mathcal{E}}^\infty$, i.e. a non-local symmetry of $\mathcal{E}$.

\section{Final comments  } 

This paper has been mainly concerned with the following two points: Jacobi multipliers and non-local symmetries. 

\begin{itemize}
\item  The  Jacobi multipliers have been first considered in relation with the inverse problem  of the Lagrangian formalism and then applied  to the study of two particular nonlinear oscillators.

\item The theory of non-local symmetries  has been studied by making use of a geometric approach. We prove that the extended formalism can be a very interesting procedure for obtaining symmetries of nonlinear systems.

\item We have shown that the use of infinite-dimensional jet manifolds does not complicate the description of non-local symmetries of systems and allows us to develop a more rigorous theoretical approach. In addition, certain structures are now naturally defined.

\item In the future we aim to apply the theory of non-local symmetries  to a generalisation of the nonlinear oscillators studied in this work that contain an isotopic term. This will describe as a particular case the non-linear oscillators detailed in \cite{BEHR08} on a one-dimensional manifold.

\item Diffieties are mainly used in the study of systems of partial differential equations. Nevertheless, we aim to show that these structures may play a r\^ole also for the study of relevant systems of first-order differential equations.
\end{itemize}

\section{Acknowledgments}
Research of J. de Lucas founded by the  Polish National Science Centre grant MAESTRO under the contract number DEC-2012/06/A/ST1/00256.
 Partial financial support by research projects MTM2012-33575, MTM2011-15725-E and E24/1 (DGA)
 are acknowledged. J. de Lucas also acknowledges a stay at the University of Zaragoza supported by Gobierno de Arag\'on (FMI43/10).


\begin{thebibliography}{38}
\bibitem{Ja44a}
C. Jacobi, 
{\it Sul principio  dell'ultimo moltiplicatore, e suo uso come nuovo principio generale di
  meccanica},  
  Giornale Arcadico di Scienze, Lettere ed Arti \textbf {99}, 129--146 (1844)

\bibitem{Cl09} 
C. Jacobi,  A. Clebsch, and  C. Brockhardt,
{\sl Jacobi's Lectures on Dynamics}, Texts and Readings in Mathematics, Hindustan Book Agency,  2009

\bibitem{KKV04}
  P. Kersten,   I. Krasil'shchik,  and   A. Verbovetsky, 
  {\it Nonlocal constructions in the geometry of PDE}, in: {\sl Symmetry in nonlinear mathematical physics. Part 1,
  2, 3}, series Pr. Inst. Mat. Nats. Akad. Nauk Ukr. Mat.
  Zastos., 50, Part 1, Vol.~ 2,
  Nats\=\i onal. Akad. Nauk Ukra\"\i ni, \=Inst. Mat., Kiev,  2004 pp. 412--423

\bibitem{KV89}
  I.~S. Krasil'shchik and   A.~M. Vinogradov,   
  {\it Nonlocal trends in the geometry of differential equations: symmetries, conservation
  laws, and B\"acklund transformations}, 
  Acta Appl. Math. \textbf  {15}, 161--209 (1989) %
  
\bibitem{BGS11}%
 M.~S. Bruzon,   M.~L. Gandarias, and M. Senthilvelan,  
 {\it On the nonlocal symmetries of certain nonlinear oscillators and their general solution}, 
  Phys. Lett. A \textbf {375}, 2985--2987 (2011)
  %
\bibitem{Ga09}%
 M.~L. Gandarias, 
 {\it Nonlocal symmetries and reductions for some ordinary differential equations},
   Teoret. Mat. Fiz. \textbf{159},
   428--437 (2009) %
   
\bibitem{GB11}%
  M.~L.  Gandarias and   M.~S. Bruz\'on,  
  {\it Reductions for some ordinary differential equations through nonlocal symmetries},
   J. Nonlinear Math. Phys. \textbf{18}, 123--133 (2011) %
   
\bibitem{MR12}%
C. Muriel and  J.~L. Romero, 
{\it Nonlocal symmetries, telescopic vector fields and $\lambda$-symmetries of ordinary differential equations},   
 Symmetry, integrability and geometry: methods and applications
  \textbf {8},  1--21  (2012) %

\bibitem{CRS04}%
  J.~F.  Cari\~nena,   M.~F. Ra\~nada  and M. Santander,   
  {\it One-dimensional model of a quantum nonlinear harmonic oscillator},  
   Rep. Math. Phys. \textbf{ 54}, 285--293 (2004) %
   
\bibitem{BEHR08}%
 \'A. Ballesteros,   A. Enciso, F.~J.  Herranz and O. Ragnisco,
 {\it A maximally superintegrable system on an
  $n$-dimensional space of nonconstant curvature},
  Phys. D \textbf{237}, 505--509 (2008) %

\bibitem{BEHRR11}%
 \'A. Ballesteros,  A. Enciso,
    F.~J. Herranz, O. Ragnisco  and D. Riglioni,  
    {\it A new exactly solvable quantum model in $N$ dimensions},
     Phys. Lett. A \textbf{375}, 1431--1435 (2011)
  %
\bibitem{CRSS04}%
J.~F. Cari\~nena,   M.~F. Ra\~nada,   M. Santander  and   M. Senthilvelan,     
  {\it A non-linear oscillator with quasi-harmonic behaviour: two- and
  $n$-dimensional oscillators},     
  Nonlinearity \textbf{17}, 1941--1963 (2004)
   %
\bibitem{V99}%
  A.~V. Bocharov {\it et al.}  
  \emph{ Symmetries and conservation laws for differential equations of mathematical physics},\
  in: {\sl Translations of Mathematical Monographs} {\bf 
   182}, American Mathematical Society,
  Providence, RI, 1999 %

\bibitem{Vi89}%
A.~M.  Vinogradov,  
{\it Symmetries and conservation laws of partial differential equations: basic notions and results}, 
Acta Appl. Math. \textbf{15},  3--21 (1989) %
   
\bibitem{Ca07}%
D. Catalano~Ferraioli,  
{\it Nonlocal aspects of $\lambda$-symmetries and ODEs reduction},
   J. Phys. A {\bf 40}, 5479--5489 (2007) %
  
\bibitem{Krasilshchik2011}%
J. Krasil'shchik and   A. Verbovetsky,
   {\it Geometry of jet spaces and integrable systems},  arXiv:1002.0077 
   
\bibitem{NL08a}%
 M.  Nucci and   P. Leach,     
 {\it  Jacobi's Last Multiplier
  and Lagrangians for multidimensional systems},
   J. Math. Phys. \textbf{49},  073517 (2008) %
   
\bibitem{NT08b}%
M. Nucci and   K. Tamizhmani,   
  {\it Lagrangians for dissipative nonlinear oscillators: the method of Jacobi
  Last Multiplier},
   J. Nonlinear Math. Phys. \textbf {17}, 167 (2010)
  %
\bibitem{NL08c}%
M.  Nucci and   P. Leach, 
{\it The Jacobi's Last  Multiplier and its applications in mechanics},
  J. Phys. Scr. \textbf{78},  065011   (2008) %
  
\bibitem{GN15}%
G. Gubbiotti and   M. Nucci,
 {\it Quantization of  quadratic Li\ 'enard-type equations by preserving Noether symmetries}, 
 J. Nonlinear Math. Phys \textbf{422},  1235--1246 (2015) %
 
\bibitem{BGM14}%
A. Buicua,   I.~A. Garc\'ia and  S. Maza,  
{\it Multiple Hopf  bifurcation in $\mathbb{R}^3$ and inverse Jacobi multipliers},
   J. Differential Equations \textbf{256},  310--325  (2014) %
   
\bibitem{MR14}%
C. Muriel and   J.~L. Romero,   
{\it The $\lambda$-symmetry reduction method and Jacobi last multipliers},
 Commun. Nonlinear Sci. Numer. Simul. \textbf{19},  807--820 (2014) %

\bibitem{NL08b}%
 M. Nucci and  P. Leach,  
 {\it An old method of Jacobi to find Lagrangians},
   J. Nonlin. Math. Phys \textbf{16}, 431--441 (2009) %

\bibitem{NT08a}%
M. Nucci and K.Tamizhmani, 
{\it Using an old  method of Jacobi to derive Lagrangians: a nonlinear dynamical system with
  variable coefficients},    
 Nuovo Cim. \textbf{125 B}, 255--269 (2010)
  %
  
\bibitem{MR09}%
 C. Muriel and   J.~L.  Romero,    
  {\it First integrals, integrating factors and $\lambda$-symmetries of second-order differential
  equations},  J. Phys. A \textbf{42},\  365207 (2009)
  %
\bibitem{CGK}%
A.~G. Choudhury, P. Guha and  B. Khanra,
{\it On the Jacobi last
  multiplier, integrating factors and the Lagrangian formulation of
  differential equations of the Painlev\'e-Gambier classification},  
   J. Math. Anal. Appl. \textbf{360},
   651--664 (2009) %
\bibitem{CGR09}%
J. Cari\~nena,   P. Guha, and  M. Ra\~nada, 
{\it A geometric approach to  higher-order Riccati chain: Darboux polynomials and constants of the
  motion},  J. Phys.: Conf. Ser. \textbf{175}, 012009 (2009) %

\bibitem{Le81}%
 C. Leubner, 
  {\it Inequivalent Lagrangians from constants of motion},
     Phys. Lett. A \textbf{86},  68--78 (1981)
  %
\bibitem{Lo96a}%
G. L\'opez, 
{\it One-dimensional autonomous systems and dissipative systems},
     Ann. Phys. \textbf{
  251},  372--383 (1996)
  %
\bibitem{CS66}%
   D. Currie and E. Saletan,  
   {\it $q$-equivalent
  particle Hamiltonians. The classical one-dimensional case},  
  J. Math. Phys. \textbf{7},  967--974 (1966) %
\bibitem{HH81}%
 S. Hojman and H. Harleston, 
{\it  Equivalent Lagrangians: multidimensional case},
    J. Math. Phys. \textbf{22},   1414--1419 (1981) %

\bibitem{CI83}%
  J. Cari\~nena\ and  A. Ibort, 
 {\it  Non-Noether constants of motion},
   J. Phys. A \textbf{16}, 1--7 (1983) %
  
\bibitem{ML74}%
 P.~M. Mathews and M. Lakshmanan,  
 {\it On a unique nonlinear oscillator},
  Quart. Appl. Math. \textbf{32}, 215--218 (1974) %
   
\bibitem{Olver}%
P.~J.  Olver, 
\emph{ Applications of Lie groups to differential equations},
  2nd ed., Graduate Texts in Mathematics {\bf 107}, Springer-Verlag, New
  York, 1993%
  
\bibitem {Me05}%
  S.~V. Meleshko, 
  \emph{Methods for constructing exact solutions of partial differential equations},  
  Mathematical and Analytical Techniques with Applications to Engineering,
 Springer, New York,  2005
 
\bibitem{DS89}%
D.~J.  Saunders,   
\emph{ The geometry of jet bundles}, 
Lecture  Notes in Mathematics {\bf  142}, Cambridge University
  Press, Cambridge, 1989 %

\bibitem{KS08}%
  D. Krupka and D. Saunders,
  \emph{Handbook of global analysis}, 
  Elsevier Science B.V., Amsterdam, 2008
    
\bibitem{St89}
H. Stephani, 
\emph{Differential equations}, 
Cambridge University Press,  Cambridge, 1989
\end{thebibliography}


\end{document}